\documentclass{ws-p8-50x6-00}

\begin{document}

\title{Correlations and Fluctuations - Introduction}

\author{K. Fia{\L}kowski}

\address{Marian Smoluchowski Instutute of Physics, Jagellonian University,\\
 ul. Reymonta 4, 30-059 Krakow, Poland\\E-mail: uffialko@if.uj.edu.pl}


\maketitle

\abstracts{
A rewiew of the development of formalism to describe correlations and fluctuations
in multiparticle production is presented. The milestones of history of
this development and some obvious sources of
correlations in experimental data are recalled. Some recent subjects concerning
the correlations are shortly discussed.}

\section{Introduction}
This talk\footnote{Invited talk at the XXX-th ISMD, Tihany, Hungary,
9-15 October 2000} is supposed to give an introduction for the session devoted to
theory of correlations and fluctuations. Thus the standard form should be a 
review of methods employed recently in the investigations of these phenomena.
However, such a review was already presented by Hans Eggers in the previous
session\cite{egg} and I would not dare to try to add anything to his comprehensive
presentation.
\par
Therefore I am going to review instead the methods and facts which are with
us for such a long time that most of them are often forgotten. In particular,
I would like to remind you that this is not only the jubilee 30th Symposium
on Multiparticle Dynamics (called in the seventies by a few other names), but also the
30th anniversary of the beginnings of modern history of investigations on
correlations and fluctuations in multiparticle production. Thus I shall recall
some milestones of this history (obviously this short note is not a serious
historical review, but rather a highly biased personal account of some facts). 
Then I review briefly obvious sources of
correlations in experimental data and conclude with a short discussion of some
recent developments.

\section{Highlights from the History of Correlations in Multiparticle Production}
In the fall of 1970 Mueller adapted for the processes of multiparticle production
a mathematical formalism known already well in the statistics and some branches
of physics\cite{mue}. The formulae are most transparent for inclusive distributions, i.e.
for the densities of particles in momentum space averaged over all events. One
defines the relations between the one- and many-particle densities $\rho_q$ and 
correlation functions $c_q$ by 

\begin {equation}
c_2(p_1,p_2) \equiv \rho_2(p_1,p_2)-\rho_1(p_1)\rho_1(p_2),
\label{eq:dfc2}
\end {equation}
\begin {eqnarray}
c_3(p_1,p_2,p_3) &\equiv& \rho(p_1,p_2,p_3)-\rho_1(p_1)c_2(p_2,p_3)-
\rho_1(p_2)c_2(p_1,p_3)\nonumber\\&-&\rho_1(p_3)c_2(p_1,p_2)-
\rho(p_1)\rho(p_2)\rho(p_3)
\label{eq:dfc3}
\end {eqnarray}
etc., where the densities $\rho_q$ are normalized to the factorial moments of 
multiplicity distributions
\begin {equation}
\int\rho_q(p_1,...p_q)d^3p_1...d^3p_q = <n(n-1)...(n-q+1)>\equiv <n_q>.
\label{eq:intrho}
\end {equation}
Thus 
\begin {equation}
f_2 \equiv \int c_2(p_1,p_2)d^3p_1d^3p_2=<n_2>-<n>^2\equiv D^2-<n>,
\label{eq:dff2}
\end {equation}
\begin {eqnarray}
f_3 &\equiv& \int c_3(p_1,p_2,p_3)d^3p_1d^3p_2d^3p_3=<n_3>-3<n>f_2-<n>^3
\nonumber\\&\equiv& <n_3>-3<n><n_2>+2<n>^3,
\label{eq:dff3}
\end {eqnarray}
etc. These formulae justify the standard connection between "correlations"
and "fluctuations" terms: we see that non-Poissonian fluctuations in multiplicity 
distributions (i.e., $f_q\neq 0$ for $q\neq 1$) imply non-zero correlations.
In a more recent notation, Mueller's "correlation integrals" $f_q$ are called
"factorial cumulants" and denoted by $\kappa_q$. It is often convenient to
use "scaled" quantities divided by a proper power of $<n>$, i.e.
\begin {equation}
F_q \equiv <n_q>/<n>^q,~~K_q\equiv \kappa_q/<n>^q.
\label{eq:dfFK}
\end {equation}
One distinguishes usually between the "short-range" correlations (SRC), i.e. terms in
$c_q$ which vanish for large differences between $p_i$ and integrate in 
$\kappa_q$-s to terms growing with $<n>$ slower than $<n>^q$, and the 
"long-range" correlations (LRC)
yielding $\kappa_q$-s growing as $<n>^q$. Details of such a distinction depend
usually on the particular model in which they are calculated. One of the first
profits from the efforts to identify the sources of LRC was the separation of
inelastic diffraction as the source of LRC\cite{wil}$^,$\cite{FM} which dominated
the early data.
\par
Relations between the factorial moments $<n_q>$ and factorial cumulants $\kappa_q$
for any values of $q$ may be derived from their expressions by the generating
function of the multiplicity distribution $P(n)$
\begin {equation}
G(z) \equiv \sum P(n)z^n
\label{eq:dfG}
\end {equation}
which are
\begin {equation}
<n_q>=d^qG(z)/dz^q|_{z=1},~~\kappa_q=d^q\ln G(z)/dz^q|_{z=1}.
\label{eq:dfGFK}
\end {equation}
This yields for the scaled moments and cumulants 
\begin {equation}
F_1=K_1=1,~~F_2=1+K_2,~~F_3=1+3K_2+K_3,
\label{eq:relFK}
\end {equation}
etc. Recently it was noted that the ratios of cumulants to moments $H_q=K_q/F_q$
are even more useful for testing the models, as the errors largely cancel in such
ratios and even for $q$ up to $10$ the data allow to calculate $H_q$ with 
relatively small uncertainties.
\par
The apparent sharp contrast between the "differential" quantities, as the 
densities $\rho_q$ and correlation functions $c_q$
and the "integral" quantities as the factorial moments $<n_q>$ and
cumulants $\kappa_q$ is in fact slightly artificial. We never measure really
"differential" distributions, as all the experimental histograms are integrals
over bins. On the other side, both the experimental limitations and theoretical
suggestions led to the use of integrals restricted to some part of phase-space
in the LHS of Eq.(\ref{eq:intrho}) and the following formulae.
\par
This line of investigations included the subject of forward-backward
correlations, the approximate KNO scaling\cite{KNO} defined by
\begin {equation}
P(n)\cong\frac{1}{<n>}f(\frac{n}{<n>})
\label{eq:KNOs}
\end {equation}
not only in the whole pase-space, but also in its parts defined by rapidity bins, 
and in particular the dependence of normalized
moments on energy and on the size of rapidity bin. The negative binomial 
distribution (NBD)
reinvented for multiparticle production by Giovannini and Van Hove\cite{GVH}  
\begin {equation}
P(n)=\frac{\Gamma(n+k)}{n!\Gamma(k)}(\frac{\overline n}{\overline n +k})^n
(\frac{k}{\overline n +k})^k
\label{eq:NBD}
\end {equation}
was found to describe reasonably well all the measured multiplicity distributions.
For a perfect fit $\overline n$ equals the experimental average multiplicity
$<n>$. Dependence of the parameters $\overline n$ and $k$ on the energy and bin width
allowed for important tests of various models. The valuable feature of NBD 
is that the superposition of $N$ NBD-s with parameters $\overline n,~k$ is  a 
NBD with parameters $N\overline n,~Nk$, as one may easily check using the generating function 
\begin {equation}
G_{N*NBD}(z)=[G_{NBD}^{\overline n,k}(z)]^N=[1+\frac{\overline n}{k}(1-z)]^{-kN}=
G_{NBD}^{N\overline n,Nk}(z).
\label{eq:NNBD}
\end {equation}
This shows in a particularly obvious way that superposition of $N$ independent
sources yields a multiplicity distribution with correlations reduced by a factor 
$1/N$ compared to the individual distribution, since for NBD $\kappa_2=1/k$.
\par
The next important step was the proposal to investigate scaled factorial moments
in the limit of very small rapidity bins 
\cite{BP} (averaged over the bin position)
\begin {equation}
<F_q>\equiv\frac{\sum_{i=1}^M<n_q>_i/M}{(\sum_{i=1}^M<n>_i/M)^q}.
\label{eq:FBP}
\end {equation}
This proposal, originally motivated by a search for possible "intermittency"
effect due to the phase transition, resulted indeed in the discovery of expected
power-like increase of $<F_q>$ with $M$ (or the inverse bin width). However,
the very universality of the effect makes its interpretation rather difficult. 
In any case, it allows to investigate in detail many interesting 
short-range correlation effects absent in the standard Monte Carlo generators, 
and in particular the Bose-Einstein (BE) interference effects, which seem to
dominate the observed increase of moments for hadronic collisions. 
Later, 2-D and 3-D bins in momentum space were used instead of simple
rapidity bins,
\cite{ochs}$^,$\cite{BS}, and variables guaranteeing uniform distributions 
were introduced\cite{BG}$^,$\cite{ochs2}. Most important modification,
however, was the replacement of the sum over "rectangular" bins by the "snake"
integrals of densities\cite{HPG} over phase-space regions defined by small momentum
differences, which suppressed greatly the statistical fluctuations\cite{car}. 
In particular, the ratios defined for investigating the BE effects were
rediscovered
 \begin {equation}
R(Q^2)=\frac{\int\int d^3p_1d^3p_2\rho_2(p_1,p_2)\delta(Q^2+(p_1-p_2)^2)}
{\int\int d^3p_1d^3p_2\rho_1(p_1)\rho_1(p_2)\delta(Q^2+(p_1-p_2)^2)}
\label{eq:RBE}
\end {equation}
and many interesting ideas on possible emergence of intermittency in such
ratios and other quantities were presented. Many of these and other results
may be found in more detail in a review by De Wolf, Dremin
and Kittel\cite{DWDK}.

\section{Some Obvious Sources of Correlations}
There are many well known sources of correlations in multiple production. We 
recall some of them here to remind the reader that any analysis of data which 
neglects them may be very misleading. In particular, the model parameters
fitted to the data may change significantly if these effects are "switched off". 
Two such "obvious" effects are indisputable:\\
i) resonance effects. It is known that the majority of observed pions comes
from resonance decays. This induces strong positive short-range correlations for
unlike-sign pions and weaker but non-negligible correlations for like-sign
pions.\\
ii) energy-momentum conservation. It gives negative correlations which extend
over full phase-space, but integrate to terms in $\kappa_q$-s which grow slower
than $<n>^q$ (thus they are neither typical SRC, nor LRC).\\
Other "obvious" sources are model-dependent but occur for such large classes
of models that many people regard them as evident. Let us give two examples:\\  
i) string effects. They are common for all the models which assume that
multiple production is the result of colour separation and the subsequent string 
breaking. Among others, they include effects of flavour ordering, local 
compensation of transverse momenta, long-range effects from flavours at the
string ends and the inter-string effects for multi-string events ("drag effect",
depletion in "no-string" sectors between jets).\\
i1) BE (or Fermi-Dirac) interference effects. Their very existence is certain,
but their impact on correlation data depends on extra assumptions. They induce positive very 
short-range correlations for identical pions and other mesons, and negative 
SRC for identical baryons. For the better known BE effects
one may investigate not just the standard "source size" effects, but also
spatial asymmetry of the source and higher order effects\cite{wei}.
\par
Obviously, many of these effects cannot be introduced by any analytic formulae.
Thus it is very useful to build the model ideas into Monte Carlo generators,
in which all such effects may be incorporated. Otherwise no
conclusions about the agreement or disagreement between the models and data
are fully reliable.

\section{Some Recent Developments} 
Most important development of the last decades was the possibility to investigate
the correlations in the framework of QCD, and not just in more or less ephemeral
models. This is described in widely accessible reviews\cite{DWDK} 
and we will not recall it here. However, let us mention
that recently the ideas of using infrared-safe quantities\cite{SW} seem almost
forgotten (perhaps as insufficiently ambitious?). One tries to derive
predictions for the experimentally measured quantities (as the factorial
moments and cumulants) not just in the lowest order of pQCD, but solving numerically
the equations for generating functions\cite{DLN}. One of the most interesting 
physical problems seems to be the role of QCD interference effects in the
correlation data and in particular in the multiplicity distributions. These
and other subjects are discussed in a recent review by Dremin and Gary\cite{DG}.
\par
The most lively discussed experimental data concern the BE effects. Waiting for
the flood of RHIC results one develops many methods which are already
used successfully at LEP. Again, we refer to the recent review by 
Weiner\cite{wei} for
more  details. Here let us mention only two subjects which are discussed
in this symposium: the controversy on the interference
effects between pions from the decay of two different W-s, which is still not resolved
\cite{STN} and the proposals to treat the effect in $e^+e^-$ collisions
in a completely different way than in heavy ion collisions\cite{and}.
The possibility to model the details of the effect (e.g. the source 
asymmetry\cite{smi}$^,$\cite{FW}) in MC generators is also widely debated.
\par
Summarizing, there is a lot of work to be done experimentally and theoretically
to exploit fully the potential of the correlation phenomena in multiparticle
production. The aim of this mini-review was to recall some highlights
from the history of this subject and to suggest that we should not forget the
old results, which may be as relevant today as they were in seventies. This
seems to be especially true for the BE effect, where the interpretation of the
data seems today equally difficult as it was thirty years ago.

\section*{Acknowledgments}
I would like to thank Tamas Cs\"org\H{o} and all the organizers of XXX-th
ISMD for their kind hospitality. Special thanks are due to Krzysztof Wo\'zniak,
without whom I would not be able to participate in this meeting. Helpful remarks
of Andrzej Bia{\l}as, Andrzej Kota\'nski and Romek Wit are gratefully acknowledged.
This work was partially supported by the KBN grants 2 P03B 086 14 and 2 P03B 010 15.

\end{document}